# Throughput and Delay Analysis of Wireless Random Access Networks

Lin Dai, *Member, IEEE* and Tony T. Lee, *Fellow, IEEE*

*Abstract*—This paper studies the network throughput and transport delay of a multihop wireless random access network based on a Markov renewal model of packet transportation. We show that the distribution of the source-to-destination (SD) distance plays a critical role in characterizing network performance. We establish necessary and sufficient condition on the SD distance for scalable network throughput, and address the optimal rate allocation issue with fairness and the QoS requirements taken into consideration. In respect to the end-to-end performance, the transport delay is explored in this paper along with network throughput. We characterize the transport delay by relating it to nodal queueing behavior and the SD-distance distribution; the former is a local property while the latter is a global property. In addition, we apply the large deviation theory to derive the tail distribution of transport delay. To put our theory into practical network operation, several traffic scaling laws are provided to demonstrate how network scalability can be achieved by localizing the traffic pattern, and a leaky bucket scheme at the network access is proposed for traffic shaping and flow control.

## I. INTRODUCTION

This paper focuses on the packet transportation issue that did not draw much attention until the emergence of wireless ad-hoc networks [1]. In contrast with cellular systems, nodes not only generate their own packets, but also relay others' packets in a wireless ad-hoc network. The transport capacity per node can be roughly expressed as $\theta = \lambda E[L]$, where $\lambda$ is the input rate (packet per unit time) of each node, and $L$ is the source-to-destination (SD) distance, or the number of hops. Given the fact that transport capacity is limited by local node throughput, the heavy burden of transport load without any restraints on the SD distance $L$ may eventually drag the network throughput down to zero.

The preceding issue was first addressed by Gupta and Kumar in [2], who embarked on extensive studies of the throughput of wireless networks. In the random network model proposed in [2], the key assumption is that each node randomly picks a destination node with equal probability among all nodes in the network. As a result, the number of possible destination nodes would grow linearly with respect to the SD distance $L$, and the source node is more likely to pick a far-away destination node than a close one. Under this assumption, the achievable network throughput scales as $\Theta(1/\sqrt{n \log n})$, where $n$ is the total number of nodes in the network. A less pessimistic scaling law, $\Theta(1/\sqrt{n})$, was observed in [3-4] using a similar network model. These results conclude that the network throughput will approach zero with an increasing $n$. This conclusion was further elaborated and verified in a series of follow-up papers [5-8]. More sophisticated models were also developed to incorporate node mobility or fading [9-13] -- they were obviously inspired by this rather pessimistic result and aimed at improving the network throughput. As mentioned earlier, the input rate $\lambda$ and the SD distance $L$ are intrinsic elements affecting network throughput. The key to scalable throughput lies in (i) the local access-protocol operation at each node; and (ii) the global condition on the distribution of SD distance $L$.

Compared to a wired network, the large amounts of intertwined interactions among distributed nodes make it difficult to model a dynamic wireless network to capture all traffic characteristics. Since network performance depends mainly on the interference and SD-distance distribution, a statistical modeling approach is more appropriate. In our statistical model, $n$ nodes are uniformly distributed over an area and each node can successfully deliver a packet only if there are no other concurrent transmissions within the interference range of the receiver. Furthermore, we assume that all source nodes comply with the same SD-distance distribution in selecting their destination nodes. The packets are forwarded to the next hop by following the minimum Euclidean distance route. The network throughput is defined to be the equilibrium throughput of each node with conservation of the overall traffic flow taken into account.

In light of the concerns and assumptions discussed above, a Markov renewal process is proposed to model the packet transportation behavior, which enables us to further explore the key network characteristics. Recently, the association between scalable throughput and traffic localization was exploited in [14], in which traffic is initially distributed around different local regions and aggregated hierarchically for longer-distance packet transportation. We establish the necessary and sufficient condition of scalable network throughput that echoes this locality principle of traffic pattern. Specifically, we show that scalable network throughput can be achieved by localizing the traffic pattern to limit the transport load as the number of nodes $n$ increases. Furthermore, we demonstrate that scalable network throughput is not the only concern in a large wireless network; the second moment of the SD distance should also be bounded to guarantee that the backlogged workload can be cleared with bounded delay. The optimal rate allocation issue is addressed with fairness and the QoS requirements taken into consideration.

Another frontier we explore is delay performance. Most prior studies focus on transport capacity as a performance measure. In actual networks, "transport delay" is an important practical consideration – high transport capacity obtained at the expense of excessive transport delay may not be acceptable for many applications. The study of delay distribution is, therefore, critical. The distribution of transport delay can be derived from our statistical network model in a straightforward manner. The combination of per-hop delay distribution and the scalable SD-distance distribution provides us a coherent landscape of the overall wireless network performance. In particular, we show that the necessary and sufficient condition of a bounded mean transport delay coincides with that of a scalable network throughput. Furthermore, we use large deviation theory to study the tail distribution of transport delay. The criteria of global network stability can be pinpointed by the lower and upper bounds jointly with a practical approximation developed here.

Our analysis clearly indicates that the distribution of SD distance $L$ is crucial for both delay and throughput performances of wireless networks. It was observed by Li et al. in [3] that some traffic patterns yield scalable network throughput while others may not. Based on the necessary and sufficient condition on the distribution of $L$, we suggest three kinds of traffic scaling laws and give examples to illustrate how to choose appropriate system parameters. Furthermore, the implementation of traffic shaping at access by using the leaky-bucket scheme is briefly discussed.

The remainder of this paper is organized as follows. Section II is devoted to the modeling of packet transportation, which lays the foundation of the whole paper. Sections III and IV focus on detailed analysis on network throughput and transport delay, respectively. Traffic scaling laws are discussed in Section V, and a traffic shaping scheme is also offered. Section VI concludes this paper.

## II. MARKOV RENEWAL MODEL OF PACKET TRANSPORTATION

Consider a homogeneous and ergodic wireless network with $n$ nodes uniformly distributed in an area $A$. Suppose that the node density $\sigma=n/A$ keeps constant and all nodes comply with the same distribution in selecting their destinations in an isotropic manner. Packets are forwarded to the destinations over multiple hops by following the minimum Euclidean distance route. Let $L$ be the SD distance (in unit of number of hops) of a newly generated packet, which is a random variable that takes values in the sample space $S_L=\{1, 2, \ldots, \varphi\}$. Let $\lambda(l)$ be the average input rate of the newly generated packets with $l$ SD hops at each node, and $\lambda = \sum_{l=1}^{\varphi} \lambda(l)$ is the total input rate of each node. The probability mass function of $L$ is then given by

$$f_L(l) = \begin{cases} \lambda(l)/\lambda & l \in S_L \\ 0 & l \notin S_L \end{cases}. \quad (1)$$

SD distance $L$ and per-hop delay $T$ are two pivotal factors characterizing the packet transportation behavior of each single packet. It is clear that $T$ is determined by the local access protocol. With the random-access protocol, a widely used approach in packet switching systems is adopted in [15] to model a buffered random-access network as a multi-queue single-server system. Consider the input buffer of each node as a Geo/G/1 queue with arrival rate $\theta$ packets per time slot, the moment-generating function $M_T(z)$ of per-hop delay $T$, including queueing delay and access delay, is presented in Appendix I. SD distance $L$ depends on global conditions such as routing and selection strategies of destinations. A Markov renewal process will be constructed to model packet transportation.

Assume time is slotted with integer units $t \in \{0,1,2,\ldots\}$. The transition of a tagged packet takes place only when the packet is successfully forwarded. Let $X_k$ denote the residual number of hops that the packet must travel after its $k$-th transition, and let $J_k$ denote the epoch at which the $k$-th transition occurs. The stochastic process $(X, J)=\{(X_k, J_k), k=0,1,\ldots\}$ is then a time-homogeneous discrete Markov renewal process with state space $S_L$, transition probability matrix of the embedded Markov chain $\mathbf{P}=\{p_{ij}\}$, and holding time distributions $\{G_{ij}(t)\}$, where $p_{ij}=\Pr\{X_{k+1}=j| X_k=i\}$ and $G_{ij}(t)=\Pr\{J_{k+1} - J_k \leq t| X_{k+1}=j, X_k=i\}$.

The transition probability matrix $\mathbf{P}$ is determined by the routing strategy and the probability mass function of SD distance $L$. We assume that the packet is always forwarded to the next hop by following the shortest route, and the network is saturated such that as soon as the tagged packet reaches its destination, a new packet will be injected into the network. It is easy to see that in this case, the state of the packet is always decremented by one until it reaches State 1. Then the transportation process of the tagged packet will renew at the next transition, and it will jump to State $l$ with probability $f_L(l)$ for some $l \in S_L$. The transition probability matrix $\mathbf{P}$ of the embedded Markov chain is, therefore, given by

$$\mathbf{P} = \begin{bmatrix} f_L(1) & f_L(2) & \cdots & f_L(\varphi-1) & f_L(\varphi) \\ 1 & & & & \\ & 1 & & & \\ & & \ddots & & \\ & & & 1 & \end{bmatrix} \quad (2)$$

and the limiting probabilities of the embedded Markov chain can then be obtained as

$$\pi_l = \frac{1}{E[L]} \sum_{i=l}^{\varphi} f_L(i), \quad l=1,\ldots,\varphi. \quad (3)$$

Before a transition takes place, the sojourn time in any given state $l$ is the per-hop delay $T_l$ with the probability mass function $f_T(t)$. Therefore, the holding time distributions are given by

$$G_{ij}(t) = \begin{cases} F_T(t) & i-j=1 \text{ or } i=1 \\ 0 & \text{otherwise} \end{cases} \quad (4)$$

and the mean holding time in state $l$ is

$$\tau_l = \sum_{i=1}^{\varphi} p_{li}\tau_{li} = \mathrm{E}[T], \quad l=1,\ldots,\varphi, \qquad (5)$$

where $\tau_{li} = \int_0^\infty t\, dG_{li}(t)$ is the mean holding time in the current state $l$ condition on the next state $i$.

The limiting probabilities of the Markov renewal process ($X$, $J$), $\tilde{\pi}_l$, $l=1,\ldots,\varphi$, can then be obtained from (3) and (5) as [16]:

$$\tilde{\pi}_l = \frac{\pi_l \tau_l}{\sum_{i=1}^{\varphi} \pi_i \tau_i} = \pi_l = \frac{1}{\mathrm{E}[L]} \sum_{i=l}^{\varphi} f_L(i), \qquad (6)$$

which indicates that the long-run fraction of time in any state $l$, $\tilde{\pi}_l$, is only determined by the distribution of SD distance, $f_L(l)$, due to the identical mean holding time given by (5).

The Markov renewal process ($X$, $J$) enables us to explore the various properties of packet transportation. Define $\hat{L}(t)$ as the residual number of hops of a packet at time $t$. We have

$$\hat{L}(t)=X_k, \quad J_k \leq t \leq J_{k+1}. \qquad (7)$$

It is easy to see that the stochastic process $\hat{\boldsymbol{L}} = \{\hat{L}(t), t=0,1,\ldots\}$ is a semi-regenerative process where the Markov renewal process ($X$, $J$) is embedded. Fig. 1 presents a sample path of $\hat{\boldsymbol{L}} = \{\hat{L}(t), t=0,1,\ldots\}$.

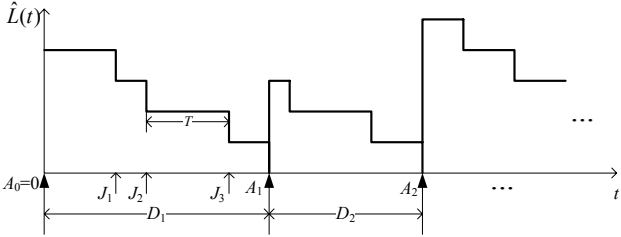

Fig. 1. A sample path of $\{\hat{L}(t), t=0,1,\ldots\}$.

The limiting probabilities of $\hat{\boldsymbol{L}}$ are given by [16]

$$\lim_{t \to \infty} \Pr\{\hat{L}(t) = l\} = \tilde{\pi}_l, \quad l=1,\ldots,\varphi. \qquad (8)$$

From (6) and (8), we immediately obtain the steady-state moment-generating function of $\hat{L}$ as follows:

$$M_{\hat{L}}(z) = \sum_{l=1}^{\infty} z^l \cdot \tilde{\pi}_l = \frac{1}{\mathrm{E}[L]} \sum_{l=1}^{\varphi} f_L(l) \sum_{k=1}^{l} z^k = \frac{z(1-M_L(z))}{(1-z)\mathrm{E}[L]}. \qquad (9)$$

Since the network is ergodic, the mean residual number of hops per packet should be the same as that of the tagged packet:

$$\mathrm{E}[\hat{L}] = \left.\frac{dM_{\hat{L}}(z)}{dz}\right|_{z=1} = \frac{\mathrm{E}[L^2]+\mathrm{E}[L]}{2\mathrm{E}[L]}. \qquad (10)$$

This is the well-known sample bias property of residual life in renewal theory [17]: The packets with larger SD distance $L$ stay in the network longer and, therefore, have a larger probability of being observed.

The sequence of regeneration epochs $\{A_0, A_1, A_2,\ldots\}$ of $\hat{\boldsymbol{L}}$ forms a renewal process, as shown in Fig. 1, where $A_i$ is the $i$th renewal of the tagged packet. The inter-arrival times, $D_i = A_i - A_{i-1}$, are i.i.d. random variables with the common probability mass function

$$\Pr\{D = x\} = \Pr\{\sum_{j=1}^{L} T_j = x\}, \quad x=1,2,\ldots \qquad (11)$$

The inter-arrival time $D$ is called the *transport delay* of the packet. The moment-generating function of $D$ can be obtained immediately from (11) as follows:

$$M_D(z) = \mathrm{E}[z^D] = \sum_{l=1}^{\infty} \mathrm{E}[z^{T_1+\cdots+T_L} | L=l] \cdot f_L(l)$$
$$= \sum_{l=1}^{\infty} M_T^l(z) \cdot f_L(l) = M_L(M_T(z)) \qquad (12)$$

where $M_T(z)$ is the moment-generating function of per-hop delay $T$. The first and second moments of transport delay $D$ are, therefore, given by

$$\mathrm{E}[D] = \mathrm{E}[L] \cdot \mathrm{E}[T] \qquad (13)$$

and

$$\mathrm{E}[D^2] = \mathrm{E}[L^2](\mathrm{E}[T])^2 + \mathrm{E}[L]\mathrm{var}[T]. \qquad (14)$$

The *residual transport delay* $\hat{D}(t)$ of a packet at time $t$ is the amount of time from $t$ to the end of the packet's journey. Let $N(t)$ be the number of renewals of the tagged packet by time $t$. The residual transport delay of a packet at time $t$, defined by $\hat{D}(t) \triangleq A_{N(t)+1} - t$, has the following limiting probabilities:

$$\lim_{t \to \infty} \Pr(\hat{D}(t) = i) = (1 - F_D(i-1))/\mathrm{E}[D], \quad i=1,2,\ldots \qquad (15)$$

The corresponding moment-generating function can be, therefore, obtained as

$$M_{\hat{D}}(z) = \frac{1}{\mathrm{E}[D]} \sum_{i=1}^{\infty} z^i (1 - F_D(i-1)) = \frac{1}{\mathrm{E}[D]} \sum_{k=1}^{\infty} f_D(k) \sum_{i=1}^{k} z^i$$
$$= \frac{z(1 - M_D(z))}{(1-z)\mathrm{E}[D]} \qquad (16)$$

and the mean residual transport delay per packet is given by

$$\mathrm{E}[\hat{D}] = \frac{\mathrm{E}[D^2] + \mathrm{E}[D]}{2\mathrm{E}[D]}. \qquad (17)$$

Combining (13) and (14), (17) can be expressed as

$$\mathrm{E}[\hat{D}] = \frac{\mathrm{E}[L^2] \cdot \mathrm{E}[T] + \mathrm{E}[L] \cdot (\mathrm{var}[T]/\mathrm{E}[T]+1)}{2\mathrm{E}[L]}. \qquad (18)$$

The performance of packet transportation is determined by SD distance $L$ and per-hop delay $T$ jointly. We will demonstrate in the subsequent sections that even though the distributions of $L$ and $T$ may vary under different system assumptions, the Markov renewal process of packet transportation ($X$, $J$) described in this section remains valid.

## III. Network Throughput

Based on the packet transportation model described in Section II, this section is devoted to a detailed analysis of network throughput. In this paper, network throughput is defined to be the equilibrium input rate $\lambda$ of each node. Let $\bar{N}$ be the total number of packets in the network. For given input rate $\lambda$ of newly generated packets at each node and mean transport delay $\mathrm{E}[D]$, the expected number of packets in the network should be equal to

$$\bar{N} = n\lambda \mathrm{E}[D] \qquad (19)$$

according to Little's Law, where $n$ is the number of nodes. On the other hand, the total arrival rate of each input buffer, which equals the node throughput $\theta$ in equilibrium, includes both the newly generated packets and the relay packets. Thus, the expected number of packets in each input buffer is given by

$\theta$E[$T$], where E[$T$] is the mean per-hop delay. Again, from Little's Law we have

$$\bar{N} = n\theta \text{E}[T]. \quad (20)$$

Given that each node has its own interference range in a wireless random-access network, a collision occurs only when concurrent transmissions are inside the same interference range. Therefore, node throughput $\theta$ should be bounded by $e^{-1}/(\pi R^2 \sigma)$, where $\pi R^2 \sigma$ is the number of nodes within the interference range of radius $R$. Combining (13), (19), and (20), we establish the following fundamental relationship between input rate $\lambda$ and local node throughput $\theta$:

$$\lambda \text{E}[L] = \theta \leq e^{-1}/(\pi R^2 \sigma). \quad (21)$$

The left side of (21) is also called the *transport capacity* of each node in [2], which includes the packets initiated by the node and the relay packets generated by other nodes. The right side of (21) is the maximum output rate per node under the random-access protocol. The inequality (21) reveals that the input rate should not exceed the maximum output rate at each node. It is a consequence of the conservation of network flows.

Note that local node throughput $\theta$ is determined by the random-access protocol, and it is a constant independent of SD distance $L$. The mean SD distance per packet E[$L$] is bounded by the maximum number of SD hops $\varphi$, which has the order of $\varphi \sim \Theta(\sqrt{n})$ in an $n$-node network. According to (21), we know that network throughput $\lambda \geq \theta/\varphi$. This indicates that the network throughput should at least have an order of $\lambda \sim \Theta(1/\sqrt{n})$. Note that it has been proven in [4] that the network throughput $\lambda$ also scales as $\lambda \sim \Theta(1/\sqrt{n})$ even with the optimal scheduling. The fact that the network throughput has the same order performance in the worst (random access) and the best (optimal scheduling) scenarios indicates that the local access protocols do not change the order results on network throughput $\lambda$.

The conservation law of network flows stated by (21) also manifests the intrinsic tradeoff between throughput $\lambda$ and mean SD distance E[$L$]. A scalable network throughput, $\lambda \sim \Theta(1)$, can only be achieved when the distribution of SD distance $L$ satisfies the constraint of bounded mean number of hops, $\lim_{n \to \infty}$E[$L$]<$\infty$. This point will be elaborated in the following subsection.

### A. Scalable Network Throughput

A scalable network throughput requires that $\lambda$ does not approach zero when the number of nodes $n$, or equivalently, the maximum number of SD hops $\varphi$, goes to infinity. Consider the special case $\sum_{l=1}^{M} l f_L(l) = 0$, which implies E[$L$] $\geq M+1$ and, therefore, $\lambda \leq \theta/(M+1)$. As a result, the network throughput will approach zero with $M$ increasing. This kind of tradeoff suggests that scalable network throughput cannot be achieved if fewer and fewer packets are delivered inside each node's proximity as the number of nodes $n$ increases.

This locality principle of traffic pattern is established in the following theorem, in which we demonstrate the necessary and sufficient condition of a scalable network throughput.

**Theorem 1**. $\lim_{\varphi \to \infty} \lambda > 0$ if and only if there exists some $M \geq 1$ such that $\lim_{\varphi \to \infty} \sum_{l=1}^{M} l \lambda(l) > 0$.

**Proof**: (1) *If*: $\lim_{\varphi \to \infty} \lambda = \lim_{\varphi \to \infty} \sum_{l=1}^{\varphi} \lambda(l) \geq \lim_{\varphi \to \infty} \sum_{l=1}^{M} \lambda(l)$
$\geq \lim_{\varphi \to \infty} \frac{1}{M} \sum_{l=1}^{M} l \lambda(l) > 0$.

(2) *Only if*: We try to prove that if for any $M \geq 1$, $\lim_{\varphi \to \infty} \sum_{l=1}^{M} l \lambda(l) = 0$, then $\lim_{\varphi \to \infty} \lambda = 0$.

For any $\varepsilon > 0$, let $a = \lceil \theta/\varepsilon \rceil$. Then $\lambda = \sum_{l=1}^{a} \lambda(l) + \sum_{l=a+1}^{\varphi} \lambda(l)$.

1) From $0 \leq \lim_{\varphi \to \infty} \sum_{l=1}^{a} \lambda(l) \leq \lim_{\varphi \to \infty} \sum_{l=1}^{a} l \lambda(l) = 0$, we know that $\lim_{\varphi \to \infty} \sum_{l=1}^{a} \lambda(l) = 0$.

2) $\sum_{l=a+1}^{\varphi} \lambda(l) = \sum_{l=a+1}^{\varphi} \frac{1}{l} \cdot l \lambda(l) \leq \theta/(a+1) < \varepsilon$. Therefore, for any $\varepsilon > 0$, there exists $a = \lceil \theta/\varepsilon \rceil$ such that $\sum_{l=a+1}^{\varphi} \lambda(l) < \varepsilon$ whenever $\varphi > a$. This means $\lim_{\varphi \to \infty} \sum_{l=a+1}^{\varphi} \lambda(l) = 0$.

By combining 1) and 2), we have $\lim_{\varphi \to \infty} \lambda = 0$. □

The criterion provided in the above theorem is equivalent to $\lim_{\varphi \to \infty}$E[$L$]<$\infty$, and both of them can be used as convenient ways to check whether a scalable throughput is achievable for a given rate allocation pattern $\lambda(l)$. For example, it is assumed in [2-8] that each node randomly picks a destination node with equal probability among all nodes in the network. This assumption leads to the following rate distribution:

$$\lambda(l) = \frac{l \lambda_0}{\varphi(\varphi+1)/2} \quad (22)$$

where $\lambda_0$ is a constant. Clearly for any $M \geq 1$, we have

$$\lim_{\varphi \to \infty} \sum_{l=1}^{M} l \lambda(l) = \lim_{\varphi \to \infty} \frac{\lambda_0 M(M+1)(2M+1)/6}{\varphi(\varphi+1)/2} = 0. \quad (23)$$

Therefore, according to Theorem 1, we conclude that no scalable network throughput can be achieved in this case.

The traffic pattern under this equal-probability-selection assumption is illustrated in Fig. 2 (a). Here the probability of the destination being a node at the periphery is always the highest because there are more nodes there. As we can see, this leads to a "vacuum zone" (i.e., the probability of selecting a destination node inside this zone is less than an arbitrary small number $\varepsilon > 0$) surrounding each source node. The radius of this vacuum zone will increase with the number of nodes $n$. In other words, in this case, the network traffic will be marginalized as the network scale increases and that will eventually drag the network throughput down to zero.

In a scalable network, the traffic should be localized so that the communication area of each node does not increase with the total number of nodes, $n$. As Fig. 2 (b) shows, Theorem 1 requires that the network traffic inside some local area does not diminish as the network grows. This usually can be guaranteed by the so-called "local preference" in practical networks. For example, it is more likely for a person in the New York City to make a telephone call to another person in the New York City than to another person in, say, Hong Kong. This does not change with network scale and is one of the principles behind organizing a practical network in a hierarchical manner. For that reason, we believe that a more realistic traffic model should take "local preference" into account.

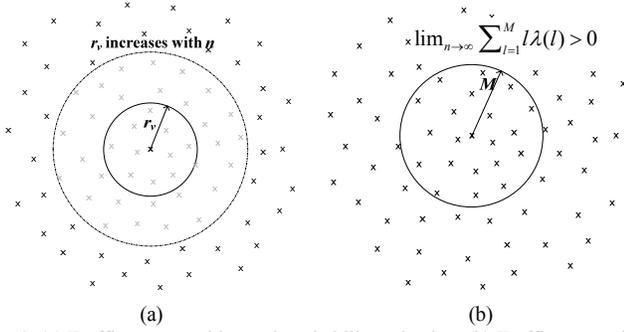

Fig. 2. (a) Traffic pattern with equal-probability-selection. (b) Traffic pattern in a scalable network.

The following examples further demonstrate the necessity of traffic locality delineated in Theorem 1.

**Example 1:** $\lambda(l)=\lambda_0/\varphi$

According to (1), the distribution of SD distance is given by $f_L(l)=1/\varphi$ and we have

$$\lim_{\varphi\to\infty} E[L] = \lim_{\varphi\to\infty}(\varphi+1)/2 = \infty. \quad (24)$$

Therefore, the network throughput is *non-scalable*.

**Example 2:** $\lambda(l)=\lambda_0\omega(l)$, in which both the probability mass function $\omega(l)$ and the maximum number of SD hops $\varphi_0$ are independent of the number of nodes $n$.

It is easy to show that the SD-distance distribution in this case $f_L(l)=\omega(l)$ implies $\lim_{\varphi\to\infty}E[L]\le\lim_{\varphi\to\infty}\varphi_0<\infty$. Therefore, the network throughput is *scalable*.

**Example 3:** $\lambda(l)=\lambda_0(1-g)^{l-1}g/l$, in which $g$ is a constant independent of the number of nodes $n$.

For any $M\ge 1$, we have

$$\lim_{\varphi\to\infty}\sum_{l=1}^{M}l\lambda(l) = \lim_{\varphi\to\infty}\theta\left(1-(1-g)^M\right) > 0. \quad (25)$$

According to Theorem 1, the network throughput is *scalable*.

### B. Optimal Rate Allocation

Efficient and fair resource allocation among SD pairs is another important issue in network management in addition to network scalability. In this subsection, the optimal rate allocation will be investigated with fairness taken into consideration.

According to (21), network throughput $\lambda$ is maximized when mean SD distance $E[L]=1$, which implies that the packets are only delivered to the nearest neighbors. In this case, the maximum network throughput $\lambda^*$ is equal to the local node throughput $\theta$.

The rate distribution $\lambda(l)$ can also be optimized with respect to the fairness of rates among different SD pairs. In particular, proportional fairness and max-min fairness are considered [20]:

i) *Proportional fairness*:

$$\max_{\lambda(l)} \sum_{l=1}^{\varphi}\log(\lambda(l)) \quad \text{subject to } \sum_{l=1}^{\varphi}l\lambda(l)=\theta. \quad (26)$$

From (26), the optimal rate allocation can be obtained as

$$\lambda_p^*(l) = \theta/(l\varphi), \ l=1,\ldots,\varphi, \quad (27)$$

which leads to a network throughput of

$$\lambda_p^* = \frac{\theta}{\varphi}\sum_{l=1}^{\varphi}1/l \approx \theta\ln\varphi/\varphi. \quad (28)$$

ii) *Max-min fairness*:

$$\max_{\lambda(l)} \min_{l=1,\ldots,\varphi} \lambda(l) \quad \text{subject to } \sum_{l=1}^{\varphi}l\lambda(l)=\theta. \quad (29)$$

It can be derived that

$$\lambda_m^*(l) = \frac{2\theta}{\varphi(\varphi+1)}, \ l=1,\ldots,\varphi, \quad (30)$$

which leads to a network throughput of

$$\lambda_m^* = 2\theta/(\varphi+1). \quad (31)$$

Comparing (28) and (31), we can see that a higher network throughput can be achieved by proportional fairness. The physical interpretation is that in the max-min fairness case, the long-distance SD pairs require more transmission resources for each unit of traffic, but yield the same input rate as the short-distance SD pairs do. They may exhaust the network resources (i.e., the airtimes required to transport the packets) and, therefore, drag down network throughput.

On the other hand, the rate distribution $\lambda(l)\sim 1/l$ optimized with respect to proportional fairness implies a window flow control scheme in which the average numbers of backlogged packets of all SD pairs are the same regardless of their SD distances. Consequently, all the SD pairs, whether long-distance or short-distance, share network resources equally, which in turn leads to a higher input rate for the short-distance SD pairs so that network throughput can be improved.

### C. Constraint on Workload Bias

The preceding discussions focus on the first moment of SD distance $L$. In this subsection, we will illustrate that the second moment of $L$ is also critical to network performance.

We have shown in Section II that the mean residual number of hops per packet $E[\hat{L}]$ represents the mean backlogged (unfinished) workload introduced by an input packet. Let $u$ denote the workload bias, which is defined as the ratio of the mean backlogged workload and the mean workload. According to (10) we have

$$u = \frac{E[\hat{L}]}{E[L]} = \frac{E[L^2]+E[L]}{2(E[L])^2} = \frac{1}{2} + \frac{1}{2}\frac{\text{var}[L]}{(E[L])^2} + \frac{1}{2E[L]}. \quad (32)$$

We can see from (32) that the workload bias $u$ is jointly determined by the first and second moments of SD distance $L$. Even though network scalability can be achieved by the bounded first moment $E[L]$ alone, an executable workload further requires bounded workload bias $u$, or equivalently, a bounded second moment $E[L^2]<\infty$.

In the example shown in Fig. 3, SD distance $L$ has a scalable first moment $\lim_{\varphi\to\infty}E[L]=2<\infty$, but a non-scalable second moment $\lim_{\varphi\to\infty}E[L^2]=\varphi=\infty$. In this case, a small fraction of packets have incredibly long SD distances and would experience unbounded delay as the number of nodes $n\to\infty$. As long as the fraction of these packets is small enough (so that $E[L]<\infty$), the network can still operate in equilibrium with a non-zero throughput. Nevertheless, these packets will stay inside the network for extremely long times and result in an unbounded backlogged workload. It is, therefore, necessary to impose constraints on both the first and second moments of SD distance $L$ so that the backlogged workload can be cleared within the delay bound.

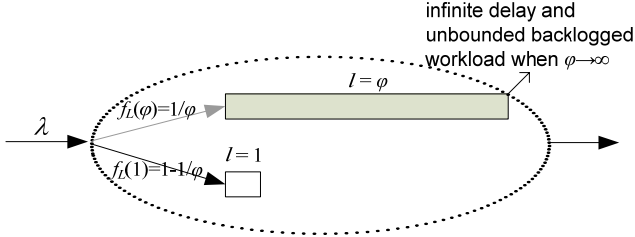

Fig. 3. When $E[L]<\infty$ and $E[L^2]=\infty$, the network can have a non-zero output but unbounded backlogged workload.

Note that the workload bias $u$ is always larger than 1/2, and it is equal to 1 when SD distance $L$ has a geometric distribution. A large $u$ indicates a large fluctuation of $L$ over the mean. According to (13) and (14), the workload bias $u$ also partially reflects the ratio of variance and square of mean of transport delay. Therefore, it can be set as a QoS requirement in shaping the distribution of SD distance $L$. In particular, by combining (21) and (32), we have

$$u = \frac{\sum_{l=1}^{\varphi} l^2 \lambda(l) \cdot \sum_{l=1}^{\varphi} \lambda(l) + \theta \cdot \sum_{l=1}^{\varphi} \lambda(l)}{2\theta^2}. \quad (33)$$

The QoS requirement on workload bias $u$ can be illustrated by the two optimal rate allocations derived in Section III.B. Substituting (27) into (33), the workload bias $u_p$ of the rate allocation subject to the proportional fairness is given by:

$$u_p = (\ln \varphi / 2 + \ln \varphi / \varphi)/2 \approx \ln \varphi / 4, \quad (34)$$

which increases with the maximum number of SD hops $\varphi$. Suppose the workload bias $u_p$ is required to be less than 1 to avoid the potentially large fluctuation of SD distance $L$ when $\varphi$ increases, then (34) indicates that the maximum number of SD hops $\varphi$ should satisfy $\varphi \leq 50$.

Similarly, the workload bias $u_m$ of the rate allocation subject to the max-min fairness can be obtained by substituting (30) into (33):

$$u_m = (2\varphi+4)/(3\varphi+3) \approx 2/3, \quad (35)$$

which is a constant less than one. This example reveals the fact that a smoother end-to-end service can actually be provided by the rate allocation scheme that yields inferior throughput.

To include the workload bias $u$ as a QoS requirement, the optimization of rate allocation can be formulated as follows:

Maximize $f(\lambda)$

Subject to $\sum_{l=1}^{\varphi} l\lambda(l) = \theta$,

and $\sum_{l=1}^{\varphi} l^2 \lambda(l) \cdot \sum_{l=1}^{\varphi} \lambda(l) + \theta \cdot \sum_{l=1}^{\varphi} \lambda(l) = 2u\theta^2$,

where $f(\lambda)$ is a suitable objective function.

## IV. TRANSPORT DELAY

The transport delay was briefly discussed in Section II in the context of the Markov renewal process. In this section, the intrinsic relationship between delay and throughput will be further explored. To demonstrate the connection between SD distance $L$ and delay performance, we use large deviation theory to study the tail distribution of transport delay.

### A. Relationship between Delay and Throughput

If we consider the entire network as a server, Little's Law states that the mean transport delay should be inversely proportional to the input rate given the mean number of packets, $E[T]\theta$, at each node. Explicitly, from (13) and (21), we have

$$E[D] = \theta E[T] / \lambda . \quad (36)$$

The mean per-hop delay $E[T]$ and local node throughput $\theta$ are determined by the local random-access protocol that does not depend on the distribution of SD distance $L$. Therefore, we can see from (36) that the necessary and sufficient condition of a bounded mean transport delay $E[D]$ is the same as that of a scalable network throughput $\lambda$ which is given in Theorem 1.

Yet another aspect of Little's Law shown in the following theorem is that the per-hop delay $T$ will become a constant if the network throughput is non-scalable.

**Theorem 2.** If $\lim_{n\to\infty} \lambda = 0$, then $T \stackrel{w.p.1}{=} E[T]$ as $n\to\infty$.

**Proof:** Let $Y=D/L=\sum_{i=1}^{L} T_i / L$. The moment generating function of $Y$ is given by

$$M_Y(z) = E[z^Y] = \sum_{l=1}^{\infty} M_T^l(z^{1/l}) \cdot f_L(l) = M_L(M_T(z^{1/l})). \quad (37)$$

Therefore,

$$E[Y] = \frac{dM_Y(z)}{dz}\bigg|_{z=1} = E[T], \quad (38)$$

$$E[Y^2] = (E[T])^2 + \text{var}[T] \cdot \sum_{l=1}^{\varphi} \frac{1}{l} f_L(l). \quad (39)$$

Combining (38) and (39), we know that

$$\text{var}[Y] = \text{var}[T] \sum_{l=1}^{\varphi} \frac{1}{l} f_L(l). \quad (40)$$

It is obvious from (40) that if $\lim_{\varphi\to\infty} \sum_{l=1}^{\varphi} \frac{1}{l} f_L(l) = 0$, then var$[Y]=0$ as $\varphi\to\infty$. Therefore, the proof can be accomplished in two steps: 1) var$[Y]=0$ implies $T \stackrel{w.p.1}{=} E[T]$, and 2) $\lim_{n\to\infty} \lambda = 0$ implies $\lim_{\varphi\to\infty} \sum_{l=1}^{\varphi} \frac{1}{l} f_L(l) = 0$.

1) According to Chebyshev's inequality, var$[Y]=0$ means $Y \stackrel{w.p.1}{=} E[Y]$, or equivalently,

$$\frac{1}{L}\sum_{i=1}^{L} T_i \stackrel{w.p.1}{=} E[T]. \quad (41)$$

Note that $L$ is a random variable, and (41) should be held for any given $l$, i.e., $\frac{1}{l}\sum_{i=1}^{l} T_i \stackrel{w.p.1}{=} E[T]$, $l=1, 2,...$. Furthermore, note that $T_i$'s are i.i.d. random variables. Let $\varepsilon > 0$ be an arbitrary small positive number. We then have

$$(\Pr(T_i > E[T]+\varepsilon))^l \leq \Pr\left(\sum_{i=1}^{l} T_i > l(E[T]+\varepsilon)\right) = 0 \quad (42)$$

for any given $l$, which implies $\Pr\{T_i > E[T]+\varepsilon\}=0$. Similarly, $\Pr\{T_i < E[T]+\varepsilon\}=0$. Therefore, $T \stackrel{w.p.1}{=} E[T]$.

2) According to (1), $\sum_{l=1}^{\varphi} \frac{1}{l} f_L(l)$ can be written as

$\sum_{l=1}^{\varphi}\frac{1}{l}\lambda(l)/\sum_{l=1}^{\varphi}\lambda(l)$. It is an easy check that both $\lim_{\varphi\to\infty}\sum_{l=1}^{\varphi}\frac{1}{l}\lambda(l)$ and $\lim_{\varphi\to\infty}\sum_{l=1}^{\varphi}\lambda(l)$ should be held under the same necessary and sufficient condition given in Theorem 1. Thus, if $\lim_{n\to\infty}\lambda=0$, then $\lim_{\varphi\to\infty}\sum_{l=1}^{\varphi}\frac{1}{l}\lambda(l)$ =0. According to L'Hospital's rule,

$$\lim_{\varphi\to\infty}\sum_{l=1}^{\varphi}\frac{1}{l}f_L(l) = \lim_{\varphi\to\infty}\left[\sum_{l=1}^{\varphi}\frac{1}{l}\lambda(l)/\sum_{l=1}^{\varphi}\lambda(l)\right] = \lim_{\varphi\to\infty}\frac{1}{\varphi} = 0 \cdot \quad (43)$$

Therefore, we conclude that if $\lim_{n\to\infty}\lambda=0$, then $T \stackrel{w.p.1}{=} \mathrm{E}[T]$ as $n\to\infty$. □

An interpretation of Theorem 2 is that there are virtually no packets in the network when the network throughput is approaching zero. Thus, the packet delay must be constant as there is neither queueing nor collision in the transmission at each node.

### B. Tail Distribution of Transport Delay

The tail distribution of transport delay $\Pr\{D>Lx\}$ is yet another important measure of the global performance of the network. We will explore the tail distribution using large deviation theory. The upper and lower bounds of $\Pr\{D>Lx\}$ are provided in Theorem 3 below.

**Theorem 3.** $M_L\left(e^{-I^-(x)}\right) \leq \Pr(D>Lx) \leq M_L\left(e^{-I^+(x)}\right)$, where $I^+(x) = \sup_{\omega}\left\{x\omega - \ln M_T\left(e^{\omega}\right)\right\}$ and $I^-(x) = -\ln\Pr\{T>x\}$.

**Proof:** According to Chernoff's formula [21], we have

$$\Pr\{T_1+\cdots+T_L > Lx \mid L=l\} \leq e^{-lI^+(x)} \quad (44)$$

where $I^+(x)$ is the rate function given by the Legendre transform of the generating function $M_T(z)$:

$$I^+(x) = \sup_{\omega}\left\{x\omega - \ln M_T\left(e^{\omega}\right)\right\}. \quad (45)$$

Therefore,

$$\Pr(D>Lx) \leq \sum_{l=1}^{\infty} e^{-lI^+(x)} f_L(l) = M_L\left(e^{-I^+(x)}\right). \quad (46)$$

On the other hand,

$$\Pr\{T_1+\cdots+T_L > Lx \mid L=l\} \geq \Pr\{T_i > x, i=1,...,L \mid L=l\}. \quad (47)$$

Therefore,

$$\Pr(D>Lx) \geq \sum_{l=1}^{\infty}\left(\Pr\{T>x\}\right)^l f_L(l) = M_L\left(\Pr\{T>x\}\right). \quad (48)$$

Substituting $\Pr\{T>x\} = e^{-I^-(x)}$ into (48), together with (46) we have $M_L(e^{-I^-(x)}) \leq \Pr(D>Lx) \leq M_L(e^{-I^+(x)})$. □

Theorem 3 reinforces that the tail distribution of transport delay is determined by the distribution of SD distance $f_L(l)$, a global property, in conjunction with the rate function, a local property. In Appendix I, we show that the moment-generating function $M_T(z)$ of the per-hop delay $T$ in a buffered Aloha network is given by

$$M_T(z) = \frac{p_0 pz[1-(1-q)z]}{(1-\theta)-[(1-pq)-p(1-q)\theta]z}, \quad (49)$$

where $p$ and $q$ are the probability of success and retransmission probability, respectively, and $p_0=1-\theta[1-p(1-q)]/(pq)$. The expressions of the corresponding rate functions $I^+(x)$ and $I^-(x)$, defined in theorem 3, are derived in Appendix II:

$$I^+(x) = -(x-1)\ln c + (x-1)\ln\phi(x) + \ln(1-\phi(x))$$
$$-\ln\left(\frac{1}{q}-\frac{1-q}{q}\cdot\frac{\phi(x)}{c}\right) + \ln\left(\frac{1}{1-c}\right) \quad (50)$$

and

$$I^-(x) = (x-1)\ln(1/c), \quad (51)$$

where $c$ and $\phi(x)$ are given in (65) and (70), respectively.

Furthermore, from Jensen's inequality, we have

$$\exp(-I^+(x)\mathrm{E}[L]) \leq M_L(e^{-I^+(x)}). \quad (52)$$

Comparing (52) with the inequality (46) given in the proof of Theorem 3, we can readily see that both $\Pr\{D>Lx\}$ and $\exp(-I^+(x)\mathrm{E}[L])$ are tight lower bounds of $M_L(e^{-I^+(x)})$. We, therefore, infer that the following approximation can be established:

$$\Pr(D>Lx) \approx \exp(-I^+(x)\mathrm{E}[L]). \quad (53)$$

This is reminiscent of Cramer's theorem stated as follows:

$$\Pr(T_1+\cdots+T_l > lx) \approx \exp(-I^+(x)l) \quad (54)$$

for a large enough $l$ [21]. The difference is that the mean SD distance $\mathrm{E}[L]$ is used in (53).

The approximation of the tail distribution given by (53) is a convenient estimation in practice because it only requires the mean value of SD distance $L$. However, the precision level depends on the distribution of $L$. Let $\Pr\{T_1+\ldots+T_L>Lx|L=l\}$ $=\exp\{-[l+\Delta(l)]I^+(x)\}$. Cramer's theorem assures that $\lim_{l\to\infty}\Delta(l)=0$, from which it can be derived that

$$\ln\Pr(D>Lx) = -I^+(x)\mathrm{E}[L](1+\delta) \quad (55)$$

where

$$\delta = \frac{1}{I^+(x)\mathrm{E}[L]}\left[I^+(x)\sum_{l=1}^{\infty}\Delta(l)f_L(l) + \sum_{k=2}^{\infty}\frac{(-1)^{k-1}}{k}\frac{\mathrm{E}\left[(\Phi-\mathrm{E}[\Phi])^k\right]}{(\mathrm{E}[\Phi])^k}\right]$$

and $\Phi=\exp\{-(L+\Delta(L))I^+(x)\}$. It is proven in Appendix III that under a certain condition, $|\delta|\sim O(1/(\mathrm{E}[L])^\alpha)$, for some $0<\alpha\leq1$. This indicates that $\exp(-I^+(x)\mathrm{E}[L])$ is a good estimation for $\Pr\{D>Lx\}$ as long as $\mathrm{E}[L]$ is sufficiently large.

To illustrate the above results, we consider a buffered Aloha network where SD distance $L$ follows the geometric distribution with parameter $\alpha_L$. Suppose the number of nodes within the interference range is $\pi R^2\sigma=10$, the retransmission probability $q=1/(\pi R^2\sigma)=0.1$ and the node throughput $\theta=0.03$. It has been shown in [15] that the probability of success $p_L$ is the stable solution of the equation $p=\exp(-\theta\pi R^2\sigma/p)$ if $q$ is chosen from the stable region. Under these assumptions, it can be derived from (49) that the mean delay $\mathrm{E}[T]=11.3$. Both the upper bound and the lower bound of the tail distribution determined by (50) and (51) are plotted in Fig. 4 along with the approximation given by (53).

Fig. 4 shows that the tail distribution will decrease as the mean SD distance $\mathrm{E}[L]$ increases. The tail distribution with a larger $\mathrm{E}[L]=100$ is significantly lower than that with a smaller

E[$L$]=5 under the same per-hop delay requirement $x$. On the other hand, it is shown in (21) that an increase in the mean SD distance E[$L$] will reduce network throughput. Therefore, the scaling parameter $\alpha_L = 1/\text{E}[L]$ should be carefully selected to strike a good balance between the network throughput and the tail distribution of transport delay. This point will be expanded in the next section.

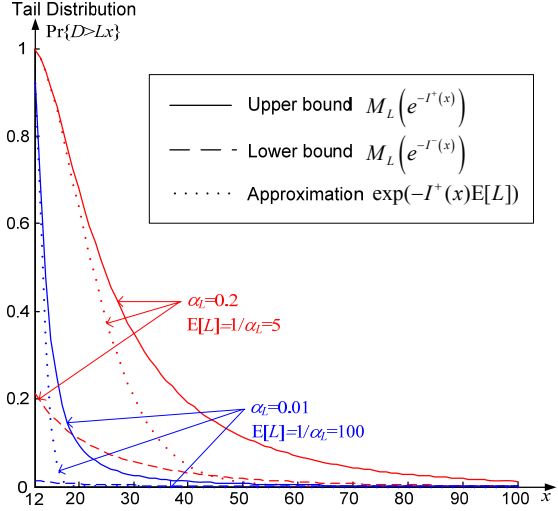

Fig. 4. Bounds and approximation of the tail distribution Pr{$D$>$Lx$} under different values of $\alpha_L$.

## V. TRAFFIC SHAPING

The previous analysis indicates that the distribution of SD distance $L$ is crucial to network performance factors such as network throughput and transport delay. In this section, we will suggest several traffic scaling laws that may shed some light on the practical network implementations in the future. A leaky-bucket scheme at the network access is also offered to show how to execute the traffic shaping.

### A. Traffic Scaling Laws

The traffic scaling law concerns the regulations of routing and access at each node to achieve scalable network throughput and bounded transport delay. We are interested in the scaling laws that govern the network traffic patterns that fulfill the conditions on the distribution of SD distance $L$ discussed in Sections III and IV.

Recall that both the first and second moments of SD distance $L$ should be bounded in a large network. To comply with these conditions, we offer three sets of traffic scaling laws to ensure that the network scalability can be accomplished by properly choosing the scaling parameters.

1) *Power Law Scaling*: $f_L(l) = c_0 l^\alpha$
2) *Exponential Scaling*: $f_L(l) = c_0 l^\alpha e^{-\beta l}$
3) *Normal Scaling*: $f_L(l) = c_0 l^\alpha e^{-\beta l^2}$

Note that SD distance $L$ is a discrete random variable representing the number of hops from the source to the destination in the previous sections. Here, for the sake of discussion, we will treat $L$ as a continuous random variable, i.e., the probability density function (pdf) of SD distance $L$, $f_L(l)$, is a continuous function in the interval [0,∞].

Let us first consider the class of power law distribution. Let $\varepsilon$>0 be the minimum SD distance. The normalization of $f_L(l)$ requires that $\int_\varepsilon^\infty f_L(l) dl = 1$, which implies $c_0 = -(1+\alpha)/\varepsilon^{1+\alpha}$ and $\alpha$<-1. A positive scaling parameter $\alpha$>0 indicates that the probability of selecting a far-away destination is always higher than that of selecting a closer one, while $\alpha$=0 refers to a uniform distribution. We know from Section III.A that neither of them can lead to a scalable network throughput. It is easy to show that if $\alpha$<-2, then a scalable throughput can be achieved because E[$L$]<∞. Moreover, a bounded second moment E[$L^2$] requires that $\alpha$<-3.

Another critical performance measure is the workload bias $u$ that reflects the ratio of variance and square of mean of transport delay. According to (32), we have

$$u = \frac{\text{E}[L^2]}{2(\text{E}[L])^2} = \frac{(\alpha+2)^2}{2(\alpha+1)(\alpha+3)}. \quad (56)$$

For a given workload bias $u$, the range of $\alpha$ will be further restricted. For example, if the workload bias $u$ is required to be less than 1, the scaling parameter $\alpha$ should satisfy $\alpha < -2 - \sqrt{2}$. It can be readily seen from (56) that a smaller $\alpha$ will lead to a lower workload bias $u$. Note that the slight difference between (56) and (32) is due to the continuous random variable assumption on SD distance $L$ that we adopted in this section.

In fact, from the cumulative distribution function (cdf) $F_L(M)=1-(M/\varepsilon)^{1+\alpha}$, we can see that the scaling parameter $\alpha$ indicates the degree of provincialism (local community interest) of the traffic pattern. For example, assume that $\varepsilon$=0.5 and $\alpha$=-10. Then approximately 99.8% packets are sent to the nearest neighbor nodes, indicating a highly localized traffic pattern. Suppose predominant traffic, for example 99%, initiated or terminated at a node is composed of packet transportations within a traffic region of radius $r_t$ that encompasses this node. It then follows from the power law distribution that the scaling parameter $\alpha$ will increase with the radius $r_t$ as follows:

$$\alpha = \frac{\log(1-99\%)}{\log(r_t/\varepsilon)} - 1. \quad (57)$$

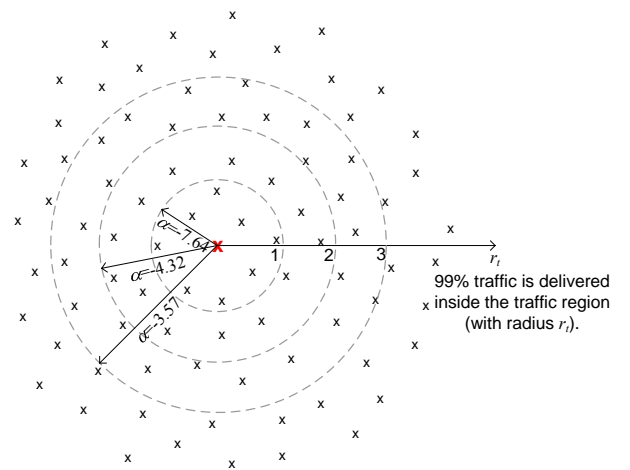

Fig. 5. Scaling factor $\alpha$ versus the radius $r_t$ of the traffic region.

As shown in Fig. 5, a smaller $\alpha$ indicates a more local traffic pattern, which leads to a higher network throughput $\lambda$ because $\lambda \sim 1+1/(1+\alpha)$ is inversely proportional to $\alpha$. To localize the traffic pattern and accommodate long-distance traffic, a hierarchical routing strategy can be adopted to keep the parameter $\alpha$ small.

If mobility is incorporated into the routing strategy, the packets will not be delivered until the nodes move into the traffic regions of their destination nodes. This routing strategy further introduces a traverse delay, which is defined as the time elapsed from when the packet was initiated until the node entered the traffic region. The traverse delay here should be differentiated from the transport delay that we have discussed in Section IV. It can be seen from Fig. 5 that a larger radius $r_t$ determined by a larger scaling parameter $\alpha$ implies that the nodes can release the packets earlier to reduce the traverse delay. However, part of the network throughput is sacrificed. Therefore, the parameter $\alpha$ can serve as a leverage to strike a balance between network throughput and traverse delay.

Our observations on the power law distribution concur with the two-phase routing strategy proposed in [9], where the source nodes distribute the packets to close-by relay nodes in the first phase, and the relay nodes deliver the packets in the second phase when they are moving into a region that is only one hop away from the destinations. It is clear that the traffic pattern of this two-hop routing strategy is highly localized. Therefore, a scalable network throughput can be guaranteed. It is, however, achieved at the cost of the large traverse delay of relay nodes that we mentioned above.

The above examples show that network performance is critically dependent on the scaling parameter. This point is reinforced by another example on how to properly choose the parameters of normal scaling law. Assume SD distance $L$ follows Rayleigh distribution, which belongs to Class 3, normal scaling, with $c_0 = 1/\sigma_L^2$, $\alpha=1$ and $\beta = 1/(2\sigma_L^2)$. According to (21), the network throughput is given by $\lambda = \theta/\mathrm{E}[L] = \theta\sqrt{2/\pi}/\sigma_L$, which indicates that the network performs better with a smaller scaling parameter $\sigma_L$.

However, this improvement on network throughput brought by a smaller $\sigma_L$ will incur a larger transport delay. The moment-generating function of $L$ is given by

$$M_L(e^t) = 1 + \sigma_L t \exp(\sigma_L^2 t^2/2)\sqrt{\pi/2}\left(\mathrm{erf}(\sigma_L t/\sqrt{2})+1\right), \quad (58)$$

which will decrease as $\sigma_L$ increases for a given $t<0$. According to Theorem 3, the tail distribution of transport delay $D$ is bounded by $M_L(e^{-I^-(x)})$ and $M_L(e^{-I^+(x)})$, which indicates that a too small $\sigma_L$ will lead to a large tail distribution of transport delay. This point can also be confirmed by (53). Suppose that $\Pr\{D>Lx\}\leq\varsigma$ is required. Let $\sigma_L^*$ be the root of equation $M_L(e^{-I^+(x)}) = \varsigma$. It is obvious that $\sigma_L^*$ is the best scaling parameter we can have, as any $\sigma_L > \sigma_L^*$ will lower the network throughput, while any $\sigma_L < \sigma_L^*$ may not satisfy the requirement on the tail distribution of transport delay.

### B. Leaky-Bucket Scheme

We have described how to choose the appropriate parameters of the proposed traffic scaling laws. This subsection is devoted to the execution of traffic shaping at the access points of the network to comply with those laws.

Fig. 6 (a) shows the leaky-bucket traffic shaping scheme implemented at each node. The newly generated input packets need to get tokens before they are transmitted. (21) indicates that both input rate $\lambda$ and SD distance $L$ contribute to the input traffic. Therefore, it requires $L$ tokens to enter the network for a packet with $L$ hops. Note that once the packet entered the network, tokens are no longer needed for its remaining journey in the network. Hence, the relay packets can go directly to the conforming buffer without getting any tokens.

The tokens are generated at a rate of $r$, which is determined by the local node throughput and bounded by $e^{-1}/(\pi R^2 \sigma)$. The bucket size $b$ is the maximum number of tokens in the bucket. Suppose $m$ input packets were conformed in time interval $[t, t+\tau]$, and there were $a$ tokens in the bucket at time $t$, $a \leq b$. We then have

$$L_1+L_2+...+L_m \leq a+r\tau \Rightarrow \frac{m}{\tau}\cdot\frac{L_1+L_2+...+L_m}{m} \leq r+\frac{a}{\tau}. \quad (59)$$

For large enough $\tau$, (59) implies $\lambda\mathrm{E}[L] \leq r+\varepsilon_0$, where $\varepsilon_0=a/\tau \leq b/\tau$ is the over-provision error.

The bucket size, $b$, is a design parameter related to the efficiency of the leaky bucket. The bucket size $b$ cannot be too large as the over-provision error is bounded by $b$. On the other hand, if $b$ is too small, excessive delay may be incurred. A compromise between error and delay is to set the bucket size between the mean and the maximum threshold of $L$, i.e., $\mathrm{E}[L]\leq b\leq L_m$, where $\Pr\{L>L_m\}\leq\varepsilon$, $\varepsilon$ is a small value. The maximum threshold $L_m$ is determined by the distribution $f_L(l)$. Take the Rayleigh distribution as an example. It can be obtained that $L_m \geq 2\sqrt{-\ln\varepsilon/\pi}\cdot\mathrm{E}[L]$. With $\varepsilon=e^{-6}<0.25\%$, $\mathrm{E}[L]\leq b\leq 2.76\mathrm{E}[L]$. Certainly, the bucket size $b$ does not have to be fixed. It can be adjusted dynamically subject to the traffic condition.

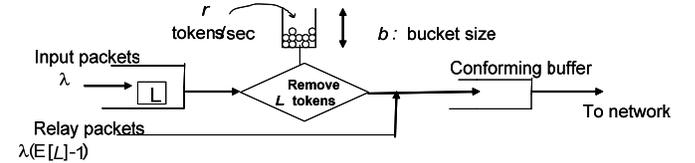

Fig. 6 (a). Leaky-bucket scheme

In the scheme described above, all new packets obtain the tokens from a common token bucket. If one of them has a large SD distance $L$, it will exhaust the token bucket and incur excessive delay for the others. To avoid such an instance, we further introduce parallel token buckets as shown in Fig. 6 (b). The new packets are assigned to different queues according to SD distance $L$, i.e., the packets with the same number of hops wait at the same queue. A token bucket is designated to each queue, and the packets can obtain tokens only from their own token buckets.

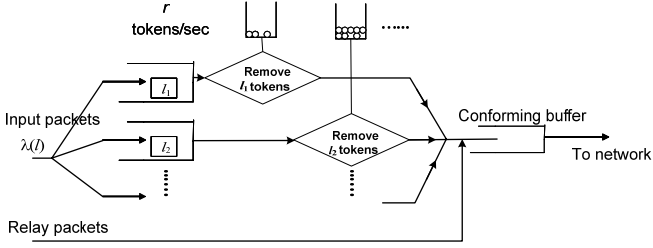

Fig. 6 (b). Leaky-bucket scheme with parallel token buckets

The tokens are still generated at a rate of $r$. However, we can adopt different rules to allocate the tokens. For example, if the tokens are equally allocated to each token bucket, the input rate of the packets with a larger SD distance $L$ will be lower, because they consume more tokens at each time. This implies a rate allocation of $\lambda(l)\sim 1/l$. On the other hand, if the tokens are allocated proportionally to SD distance $L$, the input rate of all the packets will be the same, i.e., $\lambda(l)=\lambda_0$. Clearly in the latter case, the packets with a larger $L$ occupy more network resources. As shown in Section III.B, these two rate allocation schemes actually correspond to the proportional fairness and max-min fairness, respectively. The max-min fairness case will lead to a lower network throughput.

## VI. CONCLUSIONS

In this paper, we have derived network throughput and transport delay based on a statistical wireless network model. We demonstrate that network traffic is determined by both the input rate and the distribution of SD distance. The necessary and sufficient condition for scalable network throughput shows that network scalability depends critically on how local the traffic is. We provide several traffic scaling laws and argue that a hierarchical structure would be effective to realize network scalability.

The issue of wireless networking is not just the overall throughput in general, but also which SD pairs get what throughput and the related fairness consideration. We deal with questions such as "What if all SD pairs get the same throughput regardless of their distance? And what if SD pairs of longer distance are allocated less resources? What is the impact on scaling law and what are the underlying physical interpretations?" We address these issues by formulating a resource-allocation problem and the optimal rate allocation is investigated with fairness and QoS requirements taken into consideration.

By Little's Law, the mean delay is inversely proportional to the input rate given a fixed mean number of packets in the system. Our analysis on transport delay is consistent with Little's Law, and we establish the relationship between mean transport delay and network throughput. We also develop lower and upper bounds for the tail distribution of transport delay by virtue of large deviation theory.

We conclude this paper with an illustration of how our theories may be applied in practice. We focus on the issue of traffic shaping, and show how the different parameters in leaky-bucket traffic shaping can be configured to realize the desirable characteristics as indicated by our theories. Our theories can be applied in many other ways. For example, a "window flow control" similar to that in TCP/IP networks can also be applied to ensure that long-distance SD pairs do not affect network scalability at the expense of other SD pairs. Indeed, compatibility and integration with IP networks should perhaps be a guiding goal for theories established in the realm of wireless networking. We believe and hope that our theories may provide a step toward that direction.

### APPENDIX I   MOMENT-GENERATING FUNCTION OF PER-HOP DELAY $T$

The per-hop delay $T$ represents the total waiting time of the packet, in queue and in service, at each node's buffer. For a Geo/G/1 queue with arrival rate $\theta$, the moment-generating function of $T$ can be derived as

$$M_T(z) = \frac{(1-\theta \mathrm{E}[X])(z-1)M_X(z)}{(z-1)+\theta[1-M_X(z)]} \tag{60}$$

where $X$ is the service time.

In a buffered Aloha system, a fresh head-of-line (HOL) packet will be transmitted with probability 1, and if involved in a collision, it will be retransmitted with probability $q$, until a successful transmission occurs. In this case, the service time $X$ is determined by both the probability of successful transmission $p$ and retransmission probability $q$, and the moment-generating function of $X$ can be derived as

$$M_X(z) = \frac{pz[1-(1-q)z]}{1-(1-pq)z}. \tag{61}$$

By substituting (61) into (60), we can obtain the moment-generating function of $T$ in a buffered Aloha system as

$$M_T(z) = \frac{p_0 pz[1-(1-q)z]}{(1-\theta)-[(1-pq)-p(1-q)\theta]z} \tag{62}$$

where $p_0=1-\theta[1-p(1-q)]/(pq)$. A detailed analysis on stability and throughput of buffered Aloha systems can be found in [15].

### APPENDIX II

The rate functions $I^+(x)$ and $I^-(x)$ for given $M_T(z) = \frac{p_0 pz[1-(1-q)z]}{(1-\theta)-[(1-pq)-p(1-q)\theta]z}$.

Let $\Omega(\omega)=x\omega\text{-}\ln M_T(e^\omega)$. Then

$$\frac{d\Omega}{d\omega} = x-1+\frac{(1-q)e^\omega}{1-(1-q)e^\omega}-\frac{[(1-pq)-p(1-q)\theta]e^\omega}{(1-\theta)-[(1-pq)-p(1-q)\theta]e^\omega}. \tag{63}$$

Let $y=e^\omega$, $a=1-q$, $b=[(1-pq)-p(1-q)\theta]$. (63) can be written as

$$\frac{d\Omega}{d\omega} = x-1+\frac{1}{1-ay}-1-\frac{by}{(1-\theta)-by} = x-1+\frac{1}{1-ay}-\frac{1}{1-cy} \tag{64}$$

where

$$c = b/(1-\theta) = p(1-q)+\frac{1-p}{1-\theta}. \tag{65}$$

Note that $0<q,\theta<1$. Therefore, $c>a$.

For $x>E[T]\geq 1$, Equation $d\Omega/d\omega=0$ has two roots:

$$y_1 = \frac{(c+a)+\frac{c-a}{x-1}+(c-a)\sqrt{A}}{2ac}; \quad y_2 = \frac{(c+a)+\frac{c-a}{x-1}-(c-a)\sqrt{A}}{2ac},$$

where

$$A = \left(\frac{1}{x-1}\right)^2 + 2\cdot\frac{c+a}{c-a}\frac{1}{x-1}+1. \quad (66)$$

It follows that

$$A_l = \left(\frac{1}{x-1}+1\right)^2 < A < \left(\frac{1}{x-1}+\frac{c+a}{c-a}\right)^2 = A_u. \quad (67)$$

Therefore, $y_1 > \dfrac{(c+a)+(c-a)(\frac{1}{x-1}+\sqrt{A_l})}{2ac} > \dfrac{1}{a}$, and

$$0 = \frac{(c+a)+(c-a)(\frac{1}{x-1}-\sqrt{A_u})}{2ac} < y_2 < \frac{(c+a)+(c-a)(\frac{1}{x-1}-\sqrt{A_l})}{2ac} = \frac{1}{c}.$$

From $d^2\Omega/d^2\omega<0$, we know that

$$\frac{ay}{(1-ay)^2} - \frac{cy}{(1-cy)^2} < 0 \Rightarrow (c-a)y(acy^2-1)<0. \quad (68)$$

Since only root $y_2$ satisfies (68), we have

$$\omega^* = \ln y_2 = \ln(\phi(x)/c), \quad (69)$$

where

$$\phi(x) = \frac{(c+a)+\frac{c-a}{x-1}-(c-a)\sqrt{A}}{2a}. \quad (70)$$

When $a=0$, $\phi(x)=(x-1)/x$. Clearly, $0<\phi(x)<1$. Substituting (69) into $\Omega(\omega)$, we can obtain the rate function $I^+(x)$ expressed in (50).

On the other hand, $M_T(z)$ can be written as

$$M_T(z) = \beta_1 z + \beta_2 \frac{(1-c)z}{1-cz}, \quad (71)$$

where

$$\beta_1 = \frac{p_0 p(1-q)}{c(1-\theta)} \text{ and } \beta_2 = \frac{p_0 p(c+q-1)}{c(1-c)(1-\theta)}. \quad (72)$$

We have $\Pr\{T=1\}=\beta_1+\beta_2(1-c)$ and $\Pr\{T=k\}=\beta_2(1-c)c^{k-1}$, $k>1$. By ignoring $\beta_1$, approximately we have $\Pr\{T>x\}\approx\beta_2 c^x$. According to Theorem 3, the rate function $I^-(x)$ is then given by

$$I^-(x) = -x\ln c - \ln\beta_2. \quad (73)$$

Taking $\beta_1$ into consideration, the following tighter lower bound can be obtained:

$$\Pr\{D>Lx\mid L=l\} = \sum_{m=0}^{l}\binom{l}{m}\Pr\{T_i=1, i\in\mathfrak{M}, \sum_{j\in\overline{\mathfrak{M}}}T_j>lx-m, |\mathfrak{M}|=m\}$$

$$\geq \sum_{m=0}^{l}\binom{l}{m}(\Pr(T=1))^m \Pr\{\sum_{j\in\overline{\mathfrak{M}}}T_j>lx-m, |\mathfrak{M}|=m\}$$

$$\geq \sum_{m=0}^{l}\binom{l}{m}(\beta_1+\beta_2(1-c))^m \beta_2^{l-m}c^{lx-m}$$

$$= (\beta_1+\beta_2)^l c^{l(x-1)} = c^{l(x-1)} = e^{-l(x-1)\ln(1/c)}. \quad (74)$$

Therefore, we have $I^-(x)=(x-1)\ln(1/c)$.

## APPENDIX III  PRECISION LEVEL OF APPROXIMATION
$$\Pr(D>Lx) \approx \exp(-I^+(x)E[L])$$

Suppose $Z$ is a random variable. According to Taylor's Series,

$$\ln Z = \ln a + (Z-a)/a + \sum_{k=2}^{\infty}\frac{(-1)^{k-1}}{k}\cdot\frac{(Z-a)^k}{a^k}. \quad (75)$$

Let $a=E[Z]$. From (75) we have

$$E[\ln Z] = \ln(E[Z]) + \sum_{k=2}^{\infty}\frac{(-1)^{k-1}}{k}\cdot\frac{E\left[(Z-E[Z])^k\right]}{(E[Z])^k}. \quad (76)$$

Substitute $Z$ by $\Pr\{D>Lx|L=l\}=\exp\{-(l+\Delta(l))I^+(x)\}$ in (76):

$$\ln\Pr\{D>Lx\} = -I^+(x)E[L]\cdot\left(1+\frac{1}{E[L]}\sum_{l=1}^{\infty}\Delta(l)\cdot f_L(l)\right.$$

$$\left.+\frac{1}{I^+(x)E[L]}\sum_{k=2}^{\infty}\frac{(-1)^{k-1}}{k}\cdot\frac{E\left[(\Phi-E[\Phi])^k\right]}{(E[\Phi])^k}\right) \quad (77)$$

where $\Phi=\exp\{-(L+\Delta(L))I^+(x)\}$.

Let $\delta_1 = \frac{1}{E[L]}\sum_{l=1}^{\infty}\Delta(l)f_L(l)$. We know from Cramer's theorem that $\lim_{l\to\infty}\Delta(l)=0$ and $\Delta(l)\sim o(l)$ [21]. Suppose $\Delta(l)=l^b$ for some $b<1$. $\delta_1$ can then be written as $E[L^b]/E[L]$. We consider two cases:

1) If $0<b<1$, then $E[L^b]\leq(E[L])^b$. Therefore, $0<\delta_1\leq 1/(E[L])^{1-b}$.
2) If $b<0$, then $E[L^b]\leq 1$. Therefore, $0<\delta_1\leq 1/E[L]$.

Combining the two cases 1) and 2) above, we conclude that $0<\delta_1\leq 1/(E[L])^\alpha$, for some $0<\alpha\leq 1$.

Let $\delta_2 = \dfrac{1}{I^+(x)E[L]}\sum_{k=2}^{\infty}\dfrac{(-1)^{k-1}}{k}\cdot\dfrac{E\left[(\Phi-E[\Phi])^k\right]}{(E[\Phi])^k}$. Then

$$|\delta_2| \leq \frac{1}{I^+(x)E[L]}\sum_{k=2}^{\infty}\frac{a_k}{k} \quad (78)$$

where $a_k=|E[(\Phi-E[\Phi])^k]/(E[\Phi])^k|$. If there exists a constant $\zeta$ such that $\sum_{k=2}^{\infty}a_k/k$ is bounded by $\zeta$,[1] then from (78) we know that $|\delta_2|\sim O(1/E[L])$.

Finally, we have $|\delta|\leq\delta_1+|\delta_2|\sim O(1/E[L]^\alpha)$, for some $0<\alpha\leq 1$. □

## REFERENCES

[1] A. Alwan, R. Bagrodia, N. Bambos, M. Gerla, L. Kleinrock, J. Short, and J. Villasenor, "Adaptive mobile multimedia networks," *IEEE Personal Commun. Mag.*, vol. 3, no.2, pp. 34-51, Apr. 1996.
[2] P. Gupta and P. R. Kumar, "The capacity of wireless networks," *IEEE Trans. Inf. Theory*, vol. 46, no. 2, pp. 388-404, Mar. 2000.
[3] J. Li, C. Blake, D. S. J. De Couto, H. I. Lee, and R. Morris, "Capacity of ad hoc wireless networks," in *Proc. ACM MobiCom* 2001.
[4] M. Franceschetti, O. Dousse, D. N. C. Tse, and P. Thiran, "Closing the gap in the capacity of wireless networks via percolation theory," *IEEE Trans. Inf. Theory*, vol. 53, no. 3, pp. 1009-1018, Mar. 2007.
[5] O. Dousse and P. Thiran, "Connectivity vs. capacity in dense ad hoc networks," in *Proc. Infocom* 2004.

---

[1] For example, when $f_L(l)=\exp(-l)/[\exp(-x_1)-\exp(-x_2)]$ in interval $[x_1, x_2]$ ($x_1$ is large enough so that $\Delta(l)$ can be neglected), it can be obtained that $\sum_{k=2}^{\infty}\dfrac{a_k}{k} < \sum_{k=2}^{\infty}\dfrac{1}{k(k+1)}=1/2$.

[6] A. Jovicic, P. Viswanath, and S. R. Kulkarni, "Upper bounds to transport capacity of wireless networks," *IEEE Trans. Inf. Theory*, vol. 50, no. 11, pp. 2555-2565, Nov. 2004.
[7] L.-L Xie and P. R. Kumar, "A network information theory for wireless communication: scaling laws and optimal operation," *IEEE Trans. Inf. Theory*, vol. 50, no. 5, pp. 748-767, May 2004.
[8] O. Leveque and E. Telatar, "Information-theoretic upper bounds on the capacity of large extended ad hoc wireless networks," *IEEE Trans. Inf. Theory*, vol. 51, no. 3, pp. 858-865, Mar. 2005.
[9] M. Grossglauser and D. Tse, "Mobility increases the capacity of ad hoc wireless networks," *IEEE/ACM Trans. Networking*, vol. 10, no. 4, pp. 477-486, Aug. 2002.
[10] S. Toumpis and A. J. Goldsmith, "Large wireless networks under fading, mobility and delay constraints," in *Proc. Infocom* 2004.
[11] A. E. Gamal, J. Mammen, B. Prabhakar, and D. Shah, "Throughput delay trade-off in wireless networks," in *Proc. Infocom* 2004.
[12] G. Sharma, R. Mazumdar, N. Shroff, "Delay and capacity trade-offs in mobile ad hoc networks: a global perspective," in *Proc. Infocom* 2006.
[13] F. Baccelli, B. Blaszczyszyn and P. Luhlethaler, "An Aloha protocol for multihop mobule wireless networks," *IEEE Trans. Inf. Theory*, vol. 52, no. 2, pp. 421-436, Feb. 2006.
[14] A. Ozgur, O. Leveque, and D. N. C. Tse, "Hierarchical cooperation achieves optimal capacity scaling in ad hoc networks," *IEEE Trans. Inf. Theory*, vol. 53, no. 10, pp. 3549-3572, Oct. 2007.
[15] T. T. Lee and L. Dai, "Stability and throughput of buffered Aloha with backoff," submitted to *IEEE/ACM Trans. Networking*, http://arxiv.org/abs/0804.3486 .
[16] Edward P. C. Kao, *An Introduction to Stochastic Processes*. Duxbury Press, 1996.
[17] Leonard Kleinrock, *Queueing Systems*. John Wiley & Sons, 1975.
[18] G. Bianchi, "Performance analysis of the IEEE 802.11 distributed coordination function," *IEEE J. Select. Areas Commun.*, vol. 18, no. 3, pp. 535-547, Mar. 2000.
[19] B-J Kwak, N-O Song, L. E. Miller, "Performance analysis of exponential backoff," *IEEE/ACM Trans. Networking*, pp. 343-355, April 2005.
[20] F. P. Kelly, A. K. Maulloo and D. K. H. Tan, "Rate control in communication networks: shadow prices, proportional fairness and stability," *Journal of the Operational Research Society*, vol. 49, pp. 237-252, 1998.
[21] A. Shwartz and A. Weiss, *Large Deviations for Performance Analysis – Queues, Communications and Computing*. Chapman&Hall, 1995.